\documentclass[aps,prl,twocolumn,superscriptaddress,showpacs]{revtex4-2}
\usepackage{graphicx}
\usepackage{mathrsfs}
\usepackage{bm}
\usepackage{amsmath}
\usepackage{amsfonts}
\usepackage{multirow}
\usepackage{bigstrut}
\usepackage{rotating}
\usepackage{booktabs}
\usepackage{color}
\usepackage{braket}
\usepackage{dcolumn}
\usepackage{enumerate}
\begin{document}

\title{Topological phase transition driven by magnetic field in one-dimensional topological superconductor rings}

\author{Cheng-Ming Miao}
\affiliation{College of Physics, Hebei Normal University, Shijiazhuang 050024, China}

\author{Qing-Feng Sun}
\affiliation{International Center for Quantum Materials, School of Physics, Peking University, Beijing 100871, China}
\affiliation{Collaborative Innovation Center of Quantum Matter, Beijing 100871, China}
\affiliation{CAS Center for Excellence in Topological Quantum Computation, University of Chinese Academy of Sciences, Beijing 100190, China}

\author{Ying-Tao Zhang}
\email[]{zhangyt@mail.hebtu.edu.cn}
\affiliation{College of Physics, Hebei Normal University, Shijiazhuang 050024, China}

\begin{abstract}
We study the energy spectrum and transport property of a one-dimensional
Kitaev quantum ring in a threading magnetic field. It is demonstrated that the magnetic field can effectively induce topological phase transitions for the ring
in the topologically non-trivial phase at the zero magnetic field.
However, for the ring in the topologically trivial phase at the zero field,
there is no topological phase transition, and the energy spectrum of the system is always gapped.
The magnetic field can control the appearance and disappearance of Majorana zero-energy states in the Kitaev quantum ring, when one half of the ring
is in topologically non-trivial phase and the other half is
in topologically trivial phase.
Furthermore, we calculate the transport properties of the ring connected by two semi-infinite leads. It is found that the resonant peaks of transmission coefficient $T_{\textrm{QT}}$ correspond to the critical points of topological phase transition. In addition, we extend our findings to a more realistic quantum ring adopting a semiconductor nanowire with high spin-orbit coupling, superconducting s-wave pairing and Zeeman splitting, and prove that our findings are universal.
\end{abstract}
\maketitle

\maketitle

\section{\uppercase\expandafter{\romannumeral 1}. Introduction}

Since the Majorana fermion (a particle that is its own antiparticle) is first predicted by Ettore Majorana in 1937 \cite{Majorana1937}, the hunting for it has become one of the paramount research tasks in every subfield of physics ranging from cosmology to particle physics to solid-state physics. In condensed matter physics, Alexei Kitaev \cite{Kitaev2001} suggested that the Majorana fermion could emerge as a quasiparticle excitations of Majorana zero-energy modes (MZMs) in one-dimensional spinless \emph{p}-wave superconducting chain, which is now called the Kitaev model or Kitaev chain. He also pointed out that Majorana fermions \cite{Niu2012, DeGottardi2013, Greiter2014, Hegde2015, Hegde2016} obey non-Abelian statistics \cite{Ivanov2001, Zhou2019} and are ideal candidates
for topological quantum computation \cite{Kitaev2003, Sarma2005, Nayak2008, Heck2012, Sau2011, Chen2014}. However, such a \emph{p}-wave superconducting pairing is exceedingly rare in nature. Soon Fu and Kane \cite{Fu2008} presented an ingenious proposal to artificially engineer a topological superconductor, where the traditional s-wave superconductor and topological insulator are coupled into a heterogeneous material to achieve an unpaired Majorana fermion. Lutchyn and Oreg \cite{Lutchyn2010, Oreg2010} provided an experimentally feasible setting to realize and control Majorana fermions in one-dimensional nanowires with strong spin-orbit coupling (SOC) deposited on the surface of an s-wave superconductor.

Generally, the observation of a quantized in united of $2e^{2}/h$ zero-bias conductance peak in the differential conductance measurements \cite{Mourik2012, Deng2012, Zhang2018} is considered as the experimental signatures confirming the existence of MZMs. Though several experimental groups have reported on their observations of a zero-bias conductance peak in charge transport measurements of hybrid superconductor-semiconductor nanowire devices, there is still debate for the origin of the zero-bias conductance peak. The reason is that the zero-bias conductance peaks appear even in the absence of MZMs due to disorder \cite{Liu2012, Bagrets2012, Rainis2013, DeGottardi2013a, DeGottardi2013, Adagideli2014},
weak antilocalization \cite{Pikulin2012}, nonuniform parameters \cite{Chevallier2012, Kells2012, San-Jose2013, Ojanen2013, Roy2013, Stanescu2014, Cayao2015, Fleckenstein2018} and coupling a quantum dot to a nanowire \cite{Prada2012, Liu2017}.
These peaks caused by confounding effects cannot reach the $2e^2/h$-quantized conductance plateau.
C. Moore $et. al$ \cite{Deng2018, Moore2018, Moore2018a, Liu2018, Chiu2019, Zhang2021} pointed out that even if a $2e^2/h$-quantized conductance plateau is obtained, it would not be accurate evidence for Majorana fermions.
In addition, Mao $et. al$ found that the zero-bias conductance peaks
could be changed to a zero-bias valley
with the coupling of the s-wave component of the superconductor\cite{Mao1}.
So even if no zero-bias conductance peaks are observed, it does not
mean no MZM.
Therefore, a conclusive experimental study is needed to make the unambiguous detection of MZMs.

In the previous experimental detection of topological property of the one-dimensional nanowire or two-dimensional superconductor, the researchers always focus on the experimental confirmation of topological superconductor according to the observation of MZMs in the ends of one-dimensional superconducting wires and in the
vortex cores of two-dimensional topological
superconductors \cite{Long2008,Zhang2012,Zhang2012a,He2014,Weithofer2014,Zhang2015,Doornenbal2015,Zhang2017,Danon2020,Bartolo2020,Wei2020,Li2020}, but ignore the difference of band gap in the topologically trivial and non-trivial phases. The reason is also easy to understand. No matter in the topologically trivial or non-trivial phase, there are energy gaps between the valence band and the conduction band, which cannot be directly distinguished in the experiment.

\begin{figure}
 	\centering
 	\includegraphics[width=8cm,angle=0]{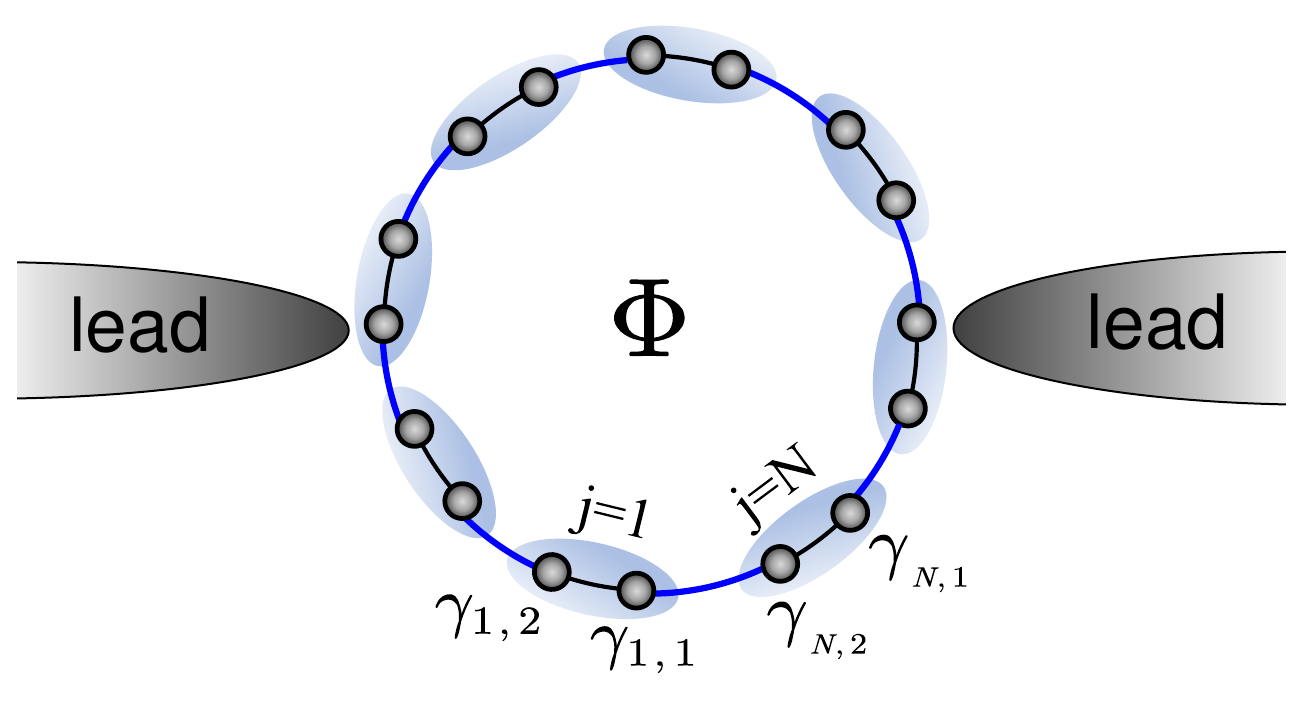}
 	\caption{Schematics of a Kitaev quantum ring with N sites connected by two semi-infinite leads. Each site $j$ hosts two Majorana modes $\gamma _{j,1}$ and $\gamma _{j,2}$. In the trivial topological phase, Majorana modes are paired within the site, linked by black arcs. In the non-trivial topological phase, Majorana modes are paired between the nearest neighbor sites, linked by blue arcs. The magnetic flux of magnitude $\Phi =N \phi$ threads through the Kitaev quantum ring.}
 	\label{fig1}
 \end{figure}

In this paper, we investigate the energy spectrum and transport property of a one-dimensional Kitaev quantum ring in a threading magnetic field, as shown in Fig. 1.  There are two Majorana modes per site, which satisfies the continuous translational symmetry in the whole quantum ring. Thus the energy spectrums whether in topologically non-trivial or trivial superconducting state will show an energy gap in the absence of a magnetic field. Interestingly, the gap of the energy spectrum in the topologically non-trivial phase closes and reopens as the strength of a magnetic field increases, accompanied by topological phase transition induced by a magnetic field like $Z_2$ topological invariants \cite{Fu2006, Fukui2007, Zeng2018}. However, the same situation does not occur in the topologically trivial phase, the gap of the energy spectrum does not change with the increasing of a magnetic field. There is no topological phase transition induced by the magnetic field in the topologically trivial phase. We also calculate the transmission coefficients as a function of the magnetic flux by employing the non-equilibrium Green's function formalism. It is demonstrated that the electron can be completely transmitted at the critical point of quantum transition induced by the magnetic field. The transport resonance peaks still correspond to the flux of the phase transition points with disorder.
Furthermore, we extend these findings to a more realistic quantum ring model by adopting a semiconductor nanowire with  high SOC, superconducting s-wave pairing and Zeeman splitting, and prove that our findings are universal.

The rest of the paper is organized as follows. In Sec. II, we present the Hamiltonian and the energy spectrum of the Kitaev quantum ring in a threading magnetic field and calculate the topological number $Z_2$ to prove the existence of Majorana states. Using the non-equilibrium Green's function, we also calculate the transmission coefficients of electron tunneling through the proposed quantum ring. In Sec. III, we present the energy spectrum and topological invariance of a more realistic quantum ring model by adopting a semiconductor nanowire with high SOC, superconducting s-wave pairing and Zeeman splitting, extend these findings as universal. Finally, a summary is presented in Sec. IV.

\section{\uppercase\expandafter{\romannumeral 2}. Kitaev quantum ring in a threading magnetic field}

We consider a one-dimensional Kitaev quantum ring in a threading magnetic field, as shown in Fig. 1, in which per-site consists of two Majorana modes and the total number of the site is chosen as $N$. The magnetic flux $\Phi=B*S$ is used to describe the magnetic field, where $B$ is the strength of the magnetic field and $S$ is the area of the quantum ring.
The Hamiltonian of the Kitaev quantum ring pierced by a magnetic flux $\Phi$ can be written as
\begin{align}
H=&\sum_{j=1}^{N}{-\mu c_{j}^{\dag}c_j}
 +\sum_{j=1}^{N-1}\left({-te^{i\phi}c_{j}^{\dag}c_{j+1}+\Delta e^{i\theta}c_{j}^{\dag}c_{j+1}^{\dag}}\right)\nonumber \\
&-te^{i\phi}c_{N}^{\dag}c_1+\Delta e^{i\theta}c_{N}^{\dag}c_{1}^{\dag}+\text{\emph{h.c.}} ,
\label{eq1}
\end{align}
where $c_{j}^{\dag} (c_{j})$ is the creation(annihilation) operator on site $j$ of the Kitaev quantum ring. The ordinary fermionic annihilation and creation operators $c_{j}$ and $c_{j}^{\dag}$ can be written in terms of Majoranas $\gamma _{j,1}$ and $\gamma _{j,2}$ by $c_j=\left( \gamma _{j,1}+i\gamma _{j,2} \right) /\sqrt{2}$, $\ c_{j}^{\dag}=\left( \gamma _{j,1}-i\gamma _{j,2} \right) /\sqrt{2}$.
$\mu$ is the chemical potential, $t$ is the nearest-neighbor hopping strength between site $j$ and $j+1$, and $\phi = \Phi/N$ is the phase of magnetic flux $\Phi$ between two nearest-neighbor sites.  $\Delta$ is the p-wave pairing amplitude and $\theta$ is the corresponding superconducting phase.
The last two terms of Eq.~(\ref{eq1}) represent the hopping and the pairing interaction between the first and last sites of the chain, which are equal to $0$ for the Kitaev chain.

\begin{figure}
	\includegraphics[width=8.5cm, clip=]{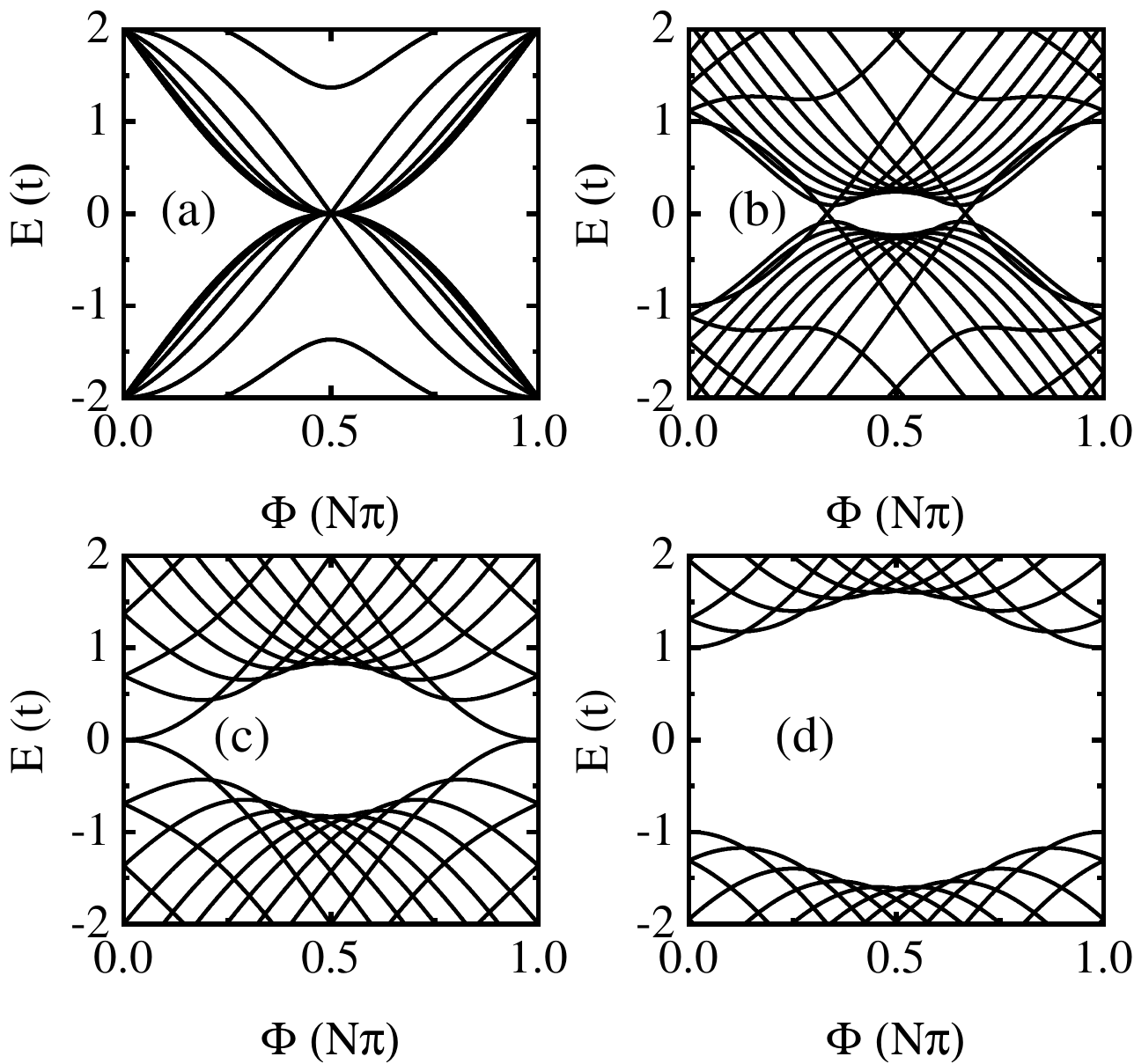}
	\caption{The energy spectrum of the Kitaev quantum ring as a function of magnetic flux $\Phi$ with different chemical potentials $\mu=0$ (a), $\mu=1$ (b), $\mu=2$ (c) and $\mu=3$ (d). The other parameters are set to be $t=1,~\Delta =1,~\theta =0,~ N=18$.}
	\label{fig2}
\end{figure}

In Fig.~\ref{fig2}, we calculate the energy spectrum of the Kitaev quantum ring as a function of magnetic flux $\Phi$ with different chemical potential $\mu$. The other parameters $\Delta$ and $t$ are set to be $1$. It has been noted that the Kitaev chain is a topologically non-trivial phase for $\left| \mu \right| < 2t$ and a topologically trivial phase
for $\left| \mu \right| > 2t$ in the absence of a magnetic filed \cite{Kitaev2001,Alicea2012}, and $\left| \mu \right| = 2t$ is the phase transition point between the two phases.
When the ring is in the topologically non-trivial phase with $\left| \mu \right| < 2t$, there is the band gap in the energy spectrum for $\Phi=0$,
as shown in Figs.~\ref{fig2}(a) and~\ref{fig2}(b).
With the increasing of the magnetic flux $\Phi$,
the gap of the energy spectrum closes and reopens at $\Phi=N\pi/2$ for $\mu=0$ in Fig.~\ref{fig2}(a) and at $\Phi=N\pi/3,~2N\pi/3$ for $\mu=1$ in Fig.~\ref{fig2}(b).
The closing and the reopening of the gap of the energy spectrum usually accompany a topological phase transition.
The middle gap region is in the topologically trivial phase,
and there are no Majorana zero-energy states at the endpoints of the Kitaev chain in open boundary conditions.
Thus the magnetic field can induce the topological phase transition in the range of $\left| \mu \right| \le 2t$.
For the phase transition point with $\left| \mu \right| = 2t$ in Fig.~\ref{fig2}(c), the gap of the energy spectrum closes at $\Phi=0$ and opens as the magnetic flux increases.
However, the energy spectrum in the topologically trivial phase
shows a completely different form.
It can be seen in Fig.~\ref{fig2}(d) that the gap of the energy spectrum always exists no matter how to change the magnetic field for $\mu=3$.
There is no topological phase transition induced
by the magnetic flux in the topologically trivial phase. Take advantage of this physical property, we can identify the system is a topologically trivial or non-trivial phase. Obviously, our proposal breaks through the limit of the zero-bias conductance platform in the open boundary condition, and directly measure the energy spectrum of the Kitaev quantum ring in the presence of the magnetic field.

It is worth mentioning that some references have studied topological superconducting rings pierced by the magnetic flux \cite{Pientka2013,Nava2017,addzou}.
For example, The Kitaev ring pierced by a magnetic flux has been studied
by F. Pientka {\sl et al.} \cite{Pientka2013} and A. Nava {\sl et al.} \cite{Nava2017}.
Usually these superconducting ring is interrupted by a weak normal link.
Then the phase of magnetic flux only appears at the weak normal link,
and the phase of magnetic flux
and the vector potential ${\bf A}$ are zero in the superconductor.
However, here we consider a complete superconducting ring
without the weak link.
The phase of magnetic flux appears in the superconductor
and every nearest-neighbor hopping term of superconducting ring
has the same phase $\phi$ of magnetic flux [see Eq.~(\ref{eq1})].
This indicates that the vector potential ${\bf A}$ in the superconductor
is not zero, and $\phi = \int_{j}^{j+1}{\bf A} \cdot d {\bf r} $.
Notice that in the superconductor, the supercurrent density
${\bf J}_s$ is proportional to the vector potential ${\bf A}$:
${\bf J}_s = \frac{e^*}{m^*}|\Psi^*|^2
\left( \hbar \nabla \theta -e^* {\bf A} \right)
=-\frac{(e^*)^2}{m^*}|\Psi^*|^2 {\bf A} $
with the mass $m^*$, charge $e^*$, and wave function $\Psi^*$ of the Cooper pair.
Here $\theta$ is the superconducting phase [see Eq.~(\ref{eq1})], which is set to zero.
So in Fig.~\ref{fig2}(b), with the increase of the magnetic flux, the vector potential
and supercurrent density of the superconducting ring increase also,
and then the topological phase transition occurs.


However, it is not sufficient for identifying the topological phase transition only considering the closing and reopening of the energy spectrum of the Kitaev quantum ring. Fortunately,  it can be seen in Fig. 1 that all unit cells in the quantum ring are equivalent by a simple translation around the ring, which indicates the translational symmetry of the ring is satisfied. Thus we can study the energy band of the system in the presence of magnetic flux. By employing Bloch's theorem we write the Hamiltonian in momentum space as
\begin{align}
H(k)&=2t\sin k\sin \phi \cdot \sigma_0-2\Delta \sin k\sin \theta \cdot \sigma_x\nonumber \\
&-2\Delta \sin k\cos \theta \cdot \sigma_y+\left( -\mu -2t\cos k\cos \phi \right) \cdot \sigma_z,
\label{eq2}
\end{align}
where $\sigma_0$ is the unit 2 $\times$ 2 matrix and $\sigma_{x,y,z}$ are the Pauli matrix representing the Numbu space.
The eigenvalue $E(k)$ is given by simple diagonalizing Eq.~(\ref{eq2})
\begin{align}
E(k) =2t\sin k\sin \phi \pm \sqrt{(\mu +2t\cos k\cos \phi)^2+4\Delta ^2\sin ^2k}.
\label{eq3}
\end{align}
One can see that the eigenvalue is independent of the superconducting phase $\theta$, as shown clearly in  Eq.~(\ref{eq3}). In our following calculations,
we set $\theta =0$.

\begin{figure}
	\includegraphics[width=8.5cm, clip=]{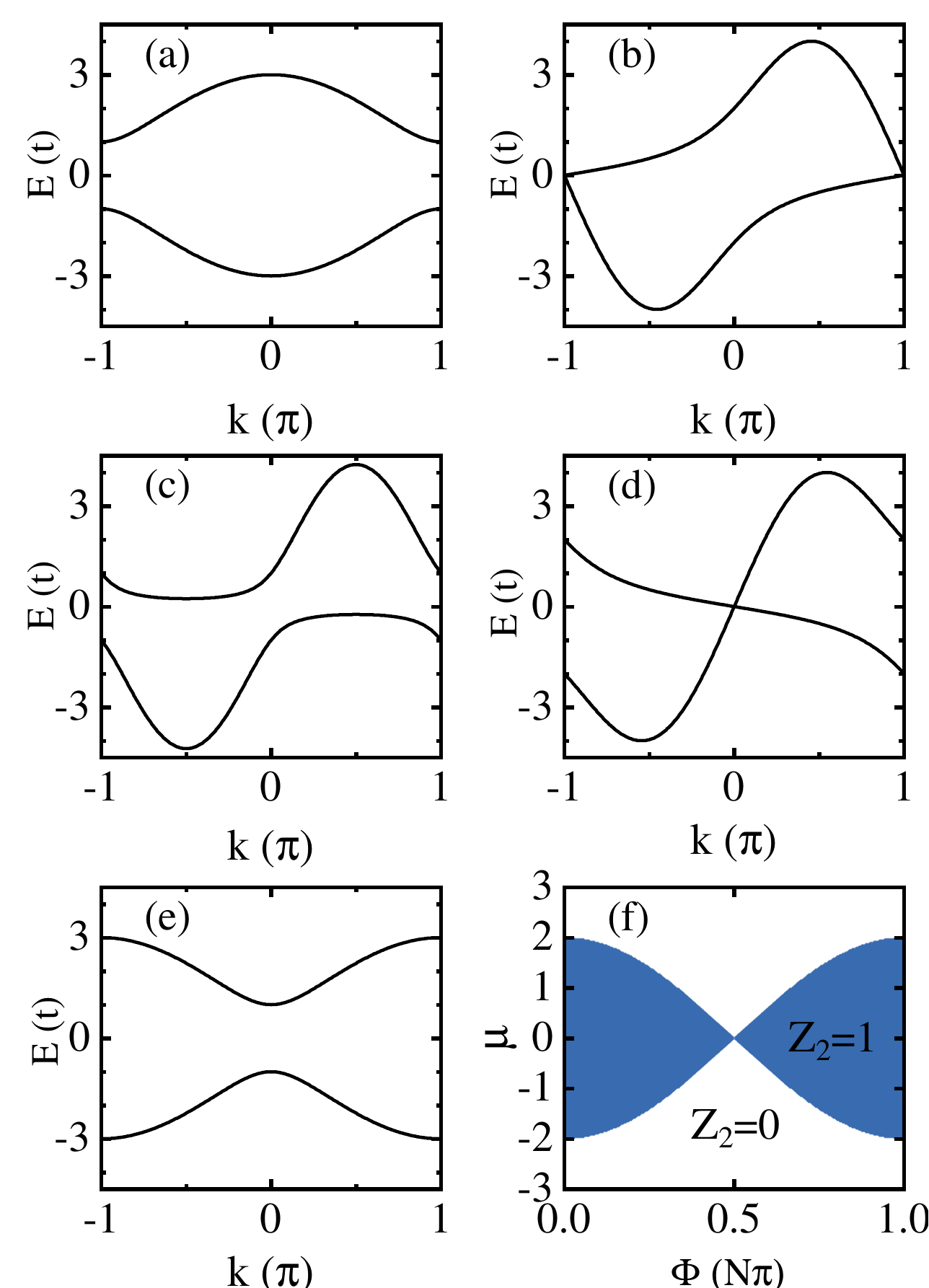}
	\caption{(a)-(e) The band structure as a function of wave-vector $k$ with different magnetic fluxes. (f) Phase diagram of the system in the plane of chemical potential $\mu$ and magnetic flux $\Phi$. The blue areas indicate that the $Z_2=1$, which are the non-trivial topology. The white areas indicate that $Z_2=0$, which are the trivial topology. The intersecting lines of the blue and white regions correspond to topological phase transition locations. The parameters are set to be $t=1,~\Delta =1,~\theta =0$, and $\mu=1$ for (a)-(e). (a) $\Phi=0$, (b) $\Phi=N\pi/3$, (c) $\Phi=N\pi/2$, (d) $\Phi=2N\pi/3$ and (e) $\Phi=N\pi$.}
	\label{fig3}
\end{figure}

We plot the band structure as a function of wave-vector $k$ with different magnetic flux in Figs.~\ref{fig3}(a) - \ref{fig3}(e).
The other parameters are set to be $t=1,~\Delta =1,~\theta =0$, and $\mu=1$ guarantees that the system is in the topologically non-trivial state in absence of a magnetic field.
The band structure has a large nontrivial band gap about $2\left| \Delta \right|$ while for the magnetic flux $\Phi = 0$, as shown in Fig. \ref{fig3}(a).
In Fig. \ref{fig3}(b) one can see that the band gap is closed at $k=\pm \pi$ for the magnetic flux $\Phi=N\pi/3$.
As the magnetic flux increases, the band gap reopens for the magnetic flux $\Phi=N\pi/2$ shown in Fig. \ref{fig3}(c).
And the band gap recloses at $k=0$ for $\Phi=2N\pi/3$ and reopens again for $\Phi=N\pi$, as plotted in Figs. \ref{fig3}(d) and \ref{fig3}(e).
It is well known that the closing and reopening of the band gap are accomplished with the topological phase transition, and the critical condition of phase transition can be obtained by solving the band-crossing condition $E(k) =2t\sin k\sin \phi \pm \sqrt{(\mu +2t\cos k\cos \phi)^2+4\Delta ^2\sin ^2k}=0$.

We further employ the Pfaffian method \cite{Fu2006} to calculate $Z_2$ topological invariants and show the phase diagram in the plane of chemical potential $\mu$ and magnetic flux $\Phi$ in Fig. \ref{fig3}(f), where the topological invariants is $Z_2=1 (0)$ blue (white) region. One can see that for $\mu=1$,  the system undergoes twice topological phase transitions from $Z_2=1$ to $Z_2=0$ and from $Z_2=0$ to $Z_2=1$ at $\Phi=N\pi/3$ and $\Phi=2N\pi/3$, respectively.
Figures \ref{fig3}(a) - \ref{fig3}(e) correspond to the five points (0,1), ($N\pi/3$,1), ($N\pi/2$,1), ($2N\pi/3$,1) and ($N\pi$,1) of the corresponding phase diagram plane ($\Phi,\mu$).
In Figs. \ref{fig3}(a) and \ref{fig3}(e) the system is topologically non-trivial phase with topological invariants $Z_2=1$ and is topologically trivial phase with $Z_2=0$ in Fig. \ref{fig3}(c), though there are the same energy gap from the energy band.
Figures \ref{fig3}(b) and \ref{fig3}(d) show the critical points of topological phase transition in the phase diagram, which is completely consistent with the conclusions obtained in Fig. \ref{fig2}(b).

\begin{figure}
	\includegraphics[width=8.5cm, clip=]{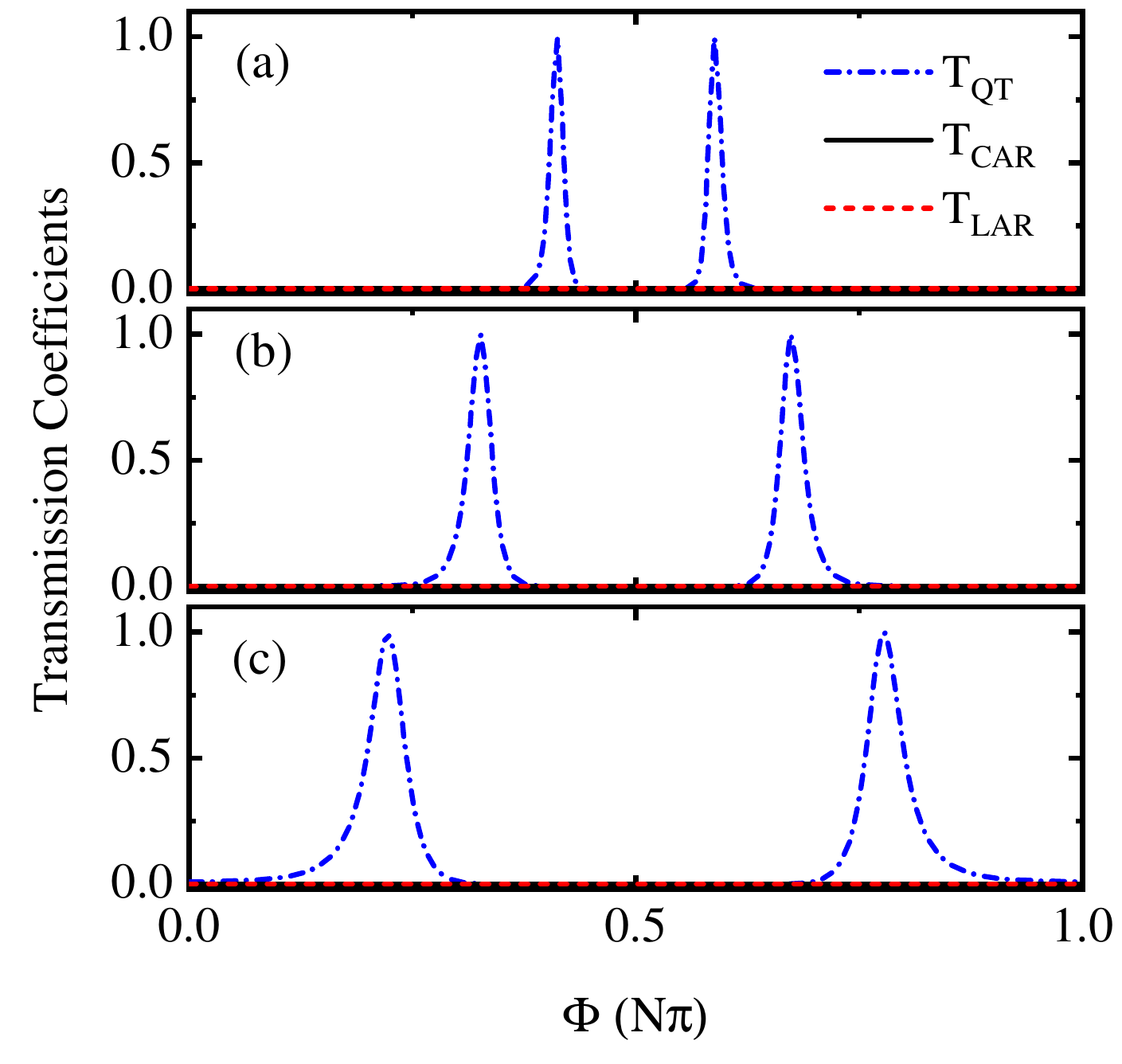}
	\caption{Transmission coefficients $T_{\textrm{QT}}$, $T_{\textrm{LAR}}$, and $T_{\textrm{CAR}}$ as a function of $\Phi$ with different chemical potentials $\mu=0.5$ for (a), $\mu=1$ for (b) and $\mu=1.5$ for (c). $E=0$, and the remaining parameters are the same as in Fig.~\ref{fig2}. The blue dotted-dashed line represents $T_{\textrm{QT}}$, the black solid line represents $T_{\textrm{CAR}}$, and the red dashed line represents $T_{\textrm{LAR}}$.}
	\label{fig4}
\end{figure}

To verify the effect of magnetic flux on the Kitaev quantum ring in more details, we employ the non-equilibrium Green's function method \cite{Sun2009} to investigate the transport property through the Kitaev quantum ring connected by two semi-infinite normal leads [see Fig. \ref{fig1}]. The leads are connected at the sites $\mathbf{j}=5$ and $\mathbf{j}=14$ in our proposal and the site number of the ring is set to be $N=18$, which guarantees the symmetry of the system.
In the process of the injected electron from left lead into the Kitaev quantum ring, there may be four possible ways: (I) the electron tunnels through the Kitaev quantum ring into the right lead as the electron, the process of which is called by the normal quantum tunneling and the coefficient of which is expressed as $T_{\text{QT}}$; (II) the electron tunnels through the Kitaev quantum ring into the right lead as the hole, the process of which is called by the crossed Andreev reflection and the coefficient of which is expressed as $T_{\text{CAR}}$;
(III) the injected electron is reflected by the Kitaev quantum ring
back to the left lead as the hole, the process of which is called by local Andreev reflection and the coefficient of which is expressed as $T_{\text{LAR}}$;
(IV) the electron is reflected by the quantum ring back to
the left lead as the electron, the process of which is called by the normal reflection. The transmission coefficients can be described as
\begin{equation}
T_{\textrm{QT}}(E)=Tr[\Gamma_{ee}^LG_{ee}^r\Gamma_{ee}^RG_{ee}^a],
\end{equation}
\begin{equation}
T_{\textrm{LAR}}(E)=Tr[\Gamma_{ee}^LG_{eh}^r\Gamma_{hh}^LG_{he}^a],
\end{equation}
\begin{equation}
T_{\textrm{CAR}}(E)=Tr[\Gamma_{ee}^LG_{eh}^r\Gamma_{hh}^RG_{he}^a],
\end{equation}
where $e$ and $h$ represent electron and hole, respectively. $\Gamma^L (E)=i[\Sigma_L^r-\Sigma_L^a]$ and $\Gamma^R (E)=i[\Sigma_R^r-\Sigma_R^a]$ are linewidth functions.
$\Sigma_L^r$ ($\Sigma_R^r$) is the self-energy due to the coupling between the left (right) lead and the central Kitaev quantum ring region, which can be numerically calculated.
$G^r(E)=[G^a(E)]^{\dagger}=[E-H_{\textrm{BdG}}-\Sigma_L^r-\Sigma_R^r]^{-1}$  are the retarded and advanced Green's functions.

In Fig. \ref{fig4}, we plot the transmission coefficients $T_{\textrm{QT}}$, $T_{\text{CAR}}$ and $T_{\text{LAR}}$ as a function of $\Phi$ with different chemical potentials $\mu$. It can be seen that the transmission coefficients $T_{\textrm{QT}}$ show two resonant peaks with the given magnetic flux period, and the location of two resonant peaks in Fig. \ref{fig4}(b) correspond to the position of closing of energy gap in Fig.~\ref{fig2}(b). Furthermore, with the increase of the chemical potential $\mu$, the width of resonant peaks is extended and the locations of resonant peaks move away from the middle symmetry axis. However, the other transmission coefficient $T_{\text{CAR}}$ and $T_{\text{LAR}}$ are suppressed and are equal to zero in the given parameters. Through the above analysis, we have known that the topological phase transition of the system can be controlled by changing the magnetic flux. Thus the process of topological phase transition can be measured by the spectral transmission properties.

\begin{figure}
	\includegraphics[width=8.5cm, clip=]{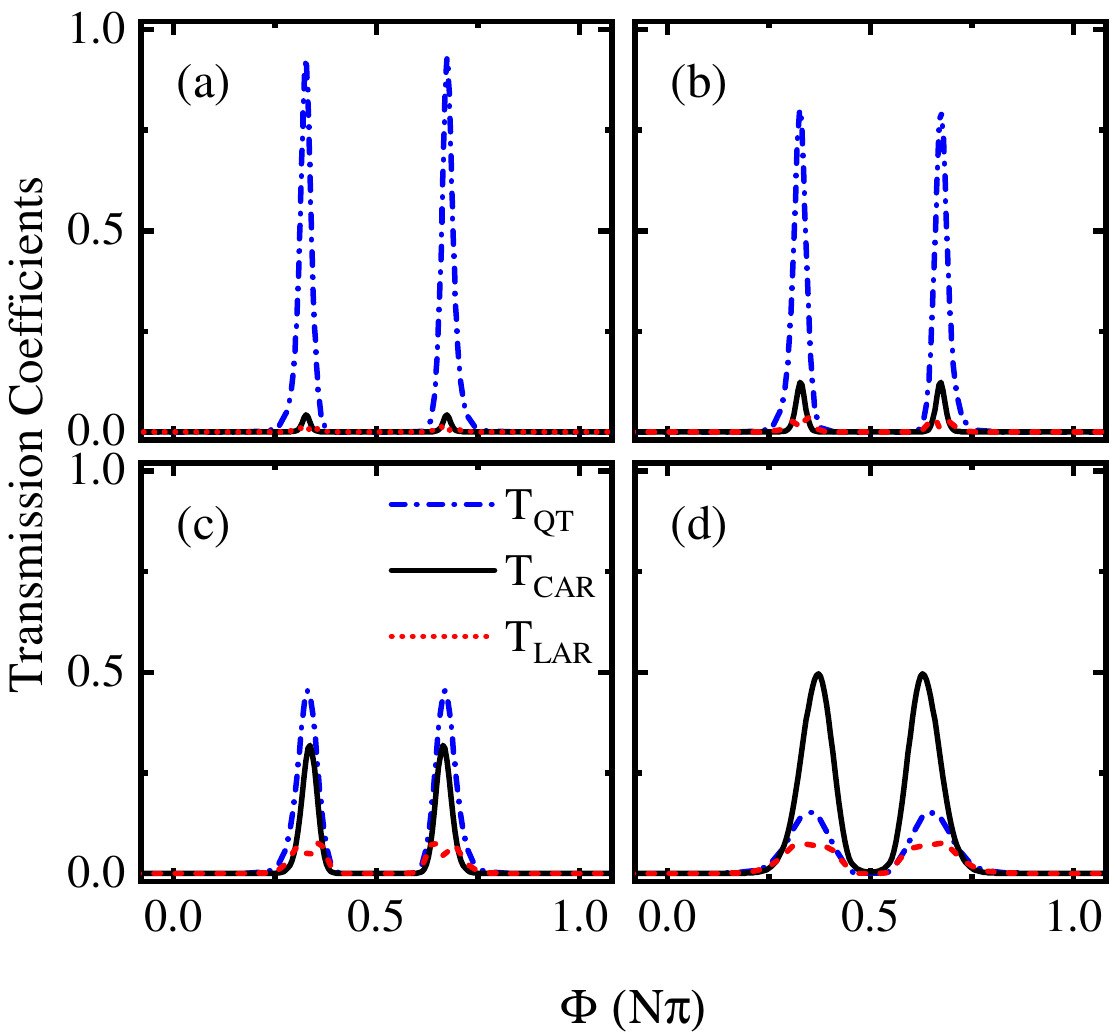}
	\caption{Transmission coefficients $T_{\textrm{QT}}$, $T_{\textrm{LAR}}$, and $T_{\textrm{CAR}}$ as a function of $\Phi$ with different disorder strength $W=0.1$ for (a), $W=0.2$ for (b), $W=0.5$ for (c) and $W=1$ for (d). $\mu=1$, and the other parameters are the same as those in Fig.~\ref{fig4}. The blue dotted-dashed line, the black solid line, and the red dashed line
	represent $T_{\textrm{QT}}$, $T_{\textrm{CAR}}$, and $T_{\textrm{LAR}}$,
	respectively.}
	\label{fig5}
\end{figure}

In Fig.~\ref{fig5}, we present the transmission coefficients $T_{\textrm{QT}}$, $T_{\text{CAR}}$ and $T_{\text{LAR}}$ as a function of $\Phi$ with different disorder strength $W$. When in the presence of the disorder,
the disorder term $\sum_{j=1}^N w_j c_j^{\dagger}c_j$ is added to Hamiltonian
in Eq.~(\ref{eq1}) with $w_j$ being uniformly distributed in the interval
[$-W$, $W$]. All the curves are averaged over 1000 random configurations, which is enough to obtain reasonable results.
From Fig.~\ref{fig5}, one can see that the two resonant peaks of the transmission coefficients $T_{\textrm{QT}}$ ($T_{\text{CAR}}$) are no longer equal to 1 (0) at the disorder strength $W=0.1$. With the increasing of the disorder strength $W$,
the peaks of $T_{\textrm{QT}}$ are suppressed and
the peaks of $T_{\text{CAR}}$ and $T_{\text{LAR}}$ gradually appear.
Even so, the peaks of $T_{\textrm{QT}}$
are still apparent at $W=0.5$ in Fig.~\ref{fig5}(c),
this indicates the resonant peaks are reliable and stable against the effects of disorder. In the presence of disorder strength $W=1$, $T_{\text{CAR}}$ replaces $T_{\textrm{QT}}$ as a dominant way [see Fig.~\ref{fig5}(d)].
It is worth noting that even if the disorder is very large,
the peaks still correspond to the flux of the phase transition point
in Fig.~\ref{fig3}(f).\\

\begin{figure}
	\includegraphics[width=8.5cm, clip=]{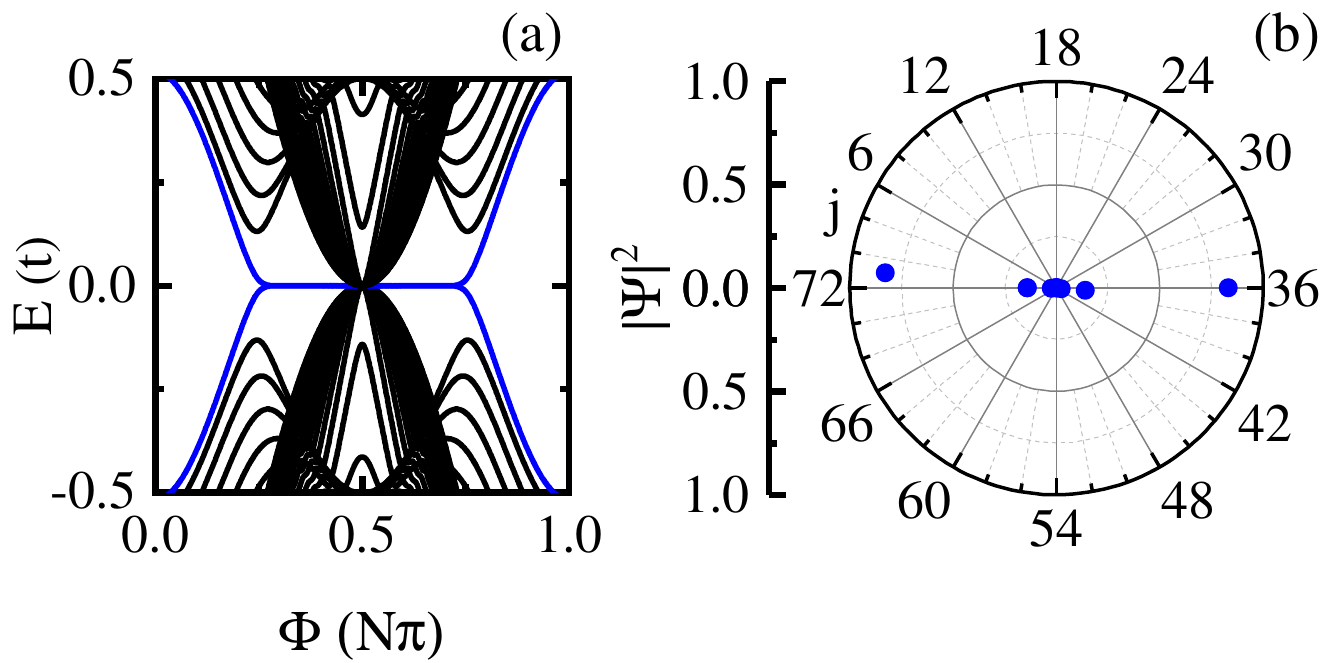}
	\caption{(a) The energy spectrum of the Kitaev quantum ring as a function of magnetic flux $\Phi$ with $\mu=0$ for 1 to 36 sites and $\mu=1.5$ for 37 to 72 sites. The flat band so-called MZMs (highlighted by the blue line) emerges as a function of magnetic flux $\Phi$. (b) Polar plot of the probability distribution of wave function $\left| \Psi \right|^2$ versus the site $j$ with the energy E = 0 for $\Phi=2N\pi/5$.
The elevant parameters are $t=1,~\Delta =1,~\theta =0,~ N=72$.}
	\label{FIG6}
\end{figure}

Because the geometry of the quantum ring has no boundaries, the end states of MZMs cannot be formed in the above analysis. In Fig. \ref{FIG6}(a) we plot the energy spectrum of the Kitaev quantum ring as a function of magnetic flux $\Phi$, where the chemical potentials of one half of the ring are set to be $\mu = 0$ and the chemical potentials of the other half are set to be $\mu = 1.5$. And the total number of the site is $N=72$.
It can be seen that the energy spectrum shows an energy gap in absence of magnetic flux $\Phi=0$, the reason is that the two halves of the quantum ring are in topologically non-trivial phases even the chemical potentials are different. As the magnetic flux increases, the energy gap closes and reopens twice,
then the flat band so-called MZMs appear and disappear as shown by the solid blue line in Fig. \ref{FIG6}(a),
indicating that the MZMs can be controlled by the magnetic flux in our proposed quantum ring. It can be illustrated that from site $j=1$ to $ j=36$ of the ring is always topologically non-trivial phase with increasing of the magnetic flux,
however, the other half part of the ring is transited into the topologically trivial phase for $\Phi$ between about $N \arccos3/4$ and $N\pi-N \arccos3/4$. Thus when $\Phi$ is in this region, half of the Kitaev quantum ring is in the topologically non-trivial phase and the other half is in the topologically trivial phase, the MZMs can be produced in the interface of two different topological phases. In Fig. \ref{FIG6}(b) we plot the probability distribution of wave function for zero-energy states as a function of site. One can see that the probability distribution of the zero-energy states is mainly located at the sites $j=72,1$ and $36, 37$ which is the interface between the two types of topological phase in real space.

\section{\uppercase\expandafter{\romannumeral 3}. Proximity-induced superconducting quantum ring in spin-orbit semiconductors}
Kitaev chain is a toy model that has not been implemented in the present material. In order to demonstrate that our above-mentioned physical discovery is universal, we extend on a more realistic quantum ring model by adopting a semiconductor nanowire with strong SOC, superconducting s-wave pairing and Zeeman splitting \cite{Stanescu2013,Yu2020,Oreg2010,Lutchyn2010,He2014}.
It has been demonstrated that the Majorana quasi-particle states exist at both ends of the semiconductor nanowires under the effect of an external magnetic field, while the semiconductor nanowire (e.g., InAs \cite{Fasth2007} or InSb \cite{Nilsson2009}) with strong SOC is coupled to a common s-wave superconductor (e.g., Nb or Al) by proximity effect. The tight-binding model of the Rashba nanowire ring is given by:
\begin{align}
H&_{R}=\sum_{n=1}^N{\sum_{\sigma}{\left( t-\mu +\sigma E_z \right)a_{n\sigma}^{\dag}a_{n\sigma}+\Delta a_{n\uparrow}a_{n\downarrow}}}\nonumber \\
&+\sum_{\sigma ;n=1}^{N-1}\left({ -\frac{t}{2}e^{i\varphi}a_{n\sigma}^{\dag}a_{n+1,\sigma}-\frac{\alpha}{2}e^{i\varphi}\sigma a_{n\sigma}^{\dag}a_{n+1,\bar{\sigma}}}\right)\nonumber \\
&-\frac{t}{2}e^{i\varphi}a_{N,\sigma}^{\dag}a_{1,\sigma}-\frac{\alpha}{2}e^{i\varphi}\sigma a_{N,\sigma}^{\dag}a_{1,\bar{\sigma}}+\text{\emph{h.c.}},
\label{eq7}
\end{align}
where $a_{n\sigma}^{\dag}(a_{n\sigma})$ is the creation(annihilation) operator on site $n$ of the Rashba nanowire ring. $\sigma=\pm 1$ for spin $\uparrow$ and $\downarrow$. $\mu$ is the chemical potential, $t$ is the nearest-neighbor hopping strength between site $n$ and $n+1$, $\phi = \Phi/N$ is the phase of magnetic flux between two nearest-neighbor sites, and $\Delta$ is the amplitude of superconducting pairing.
The extra parts are the Zeeman field $E_z$ and Rashba SOC strength $\alpha$.
The last two terms of Eq.~(\ref{eq7}) represent the hopping term between the first $n=1$ and the last $n=N$ sites of the chain. The system will be returned to the quantum chain if the last two terms are equal to be $0$.

\begin{figure}
\includegraphics[width=8.5cm, clip=]{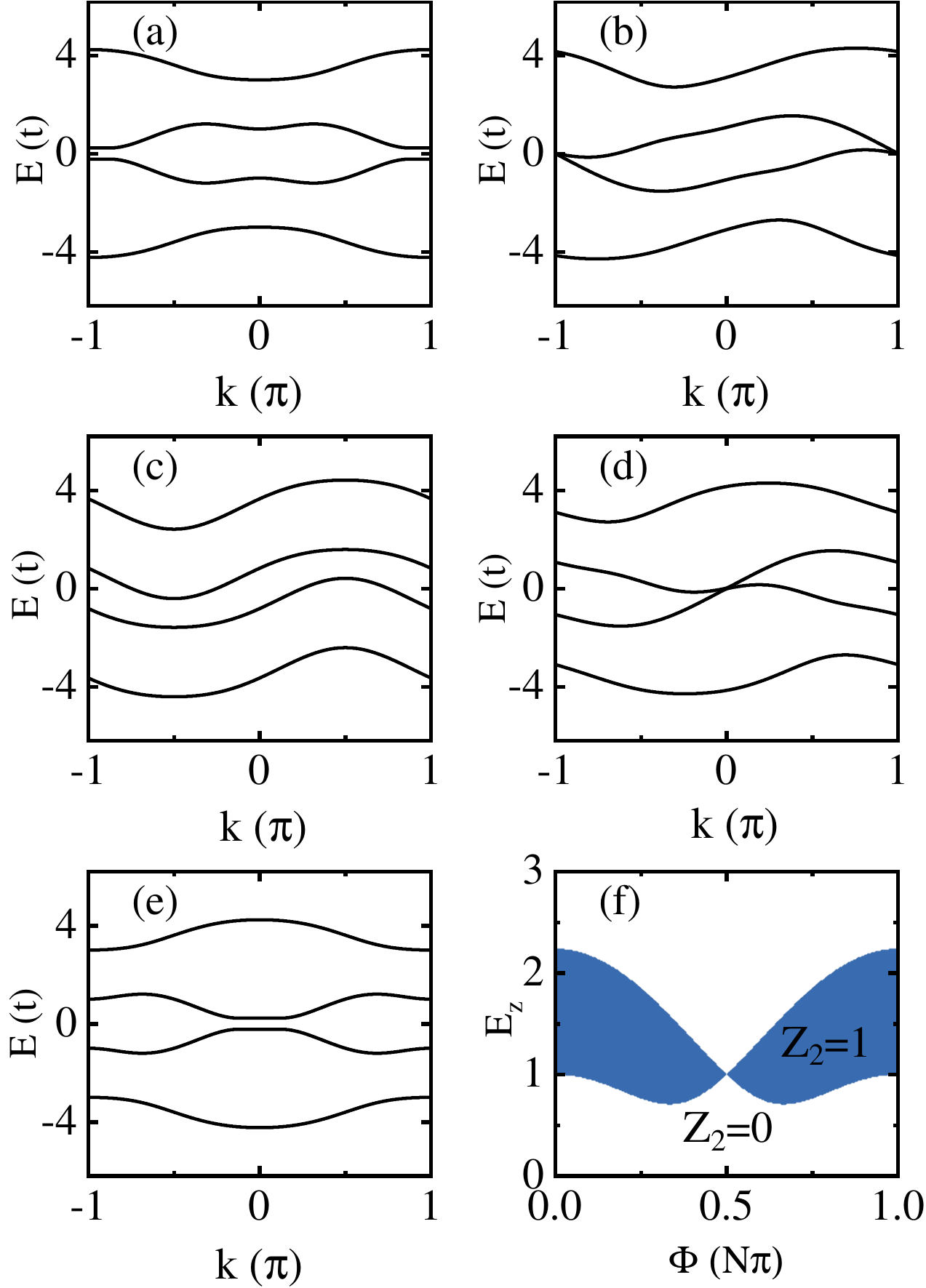}
\caption{(a)-(e) The energy band structure as a function of wave-vector $k$ with different magnetic fluxes. (f) Phase diagram of the system in the plane of Zeeman field $E_z$ and magnetic flux $\Phi$. The blue areas indicate that the $Z_2=1$, which are the non-trivial topology. The white areas indicate that $Z_2=0$, which are the trivial topology. The intersecting lines of the blue and white regions correspond to topological phase transition locations. The parameters are set to be $t=1,~\Delta =1,~\mu=0, \alpha =1$, and $E_z=2$ for (a)-(e). (a) $\Phi=0$, (b) $\Phi=N\pi/5$, (c) $\Phi=N\pi/2$, (d) $\Phi=4N\pi/5$ and (e) $\Phi=N\pi$.}
\label{FIG7}
\end{figure}

Due to the translational symmetry of the ring is satisfied, we employ Bloch's theorem to plot the band structure of the system. The topological phase transition of the system can be studied from the closing and opening of the band gap. We can transform the hamiltonian $H_{R}$ of real space into momentum space by Fourier transform as
\begin{align}
&H_{R}(k)=\left[ \left( t-\mu \right)+E_z s_z \right] \tau _z+t\sin{\varphi}k\tau _0+\alpha\cos{\varphi}ks_y\tau _z\nonumber \\
&-t\cos \varphi \left( 1-\frac{k^2}{2} \right) \tau _z+\alpha\sin \varphi \left( 1-\frac{k^2}{2} \right) s_y+\Delta s_y\tau _y,
\label{eq8}
\end{align}
and the basis vector is $\Psi_k^{\dag}=(\psi _{k\uparrow}^{\dag},\psi _{k\downarrow}^{\dag},\psi _{-k\uparrow},\psi _{-k\downarrow})$. The Pauli matrices $s_i$ and $\tau_i$ ($i=x,y,z$) act on the spin and particle-hole space respectively.
It has been noted that this system is in a topological superconducting phase \cite{Levine2017} for $E_z^2>\mu^2+\Delta^2$ in the absence of a magnetic field.

We show the band structure as a function of wave-vector $k$ with different magnetic fluxes in Figs.~\ref{FIG7}(a) - \ref{FIG7}(e).
The parameters are set to be $t=1, \Delta =1, \mu=0, \alpha =1 $, and $E_z=2$ is fixed to guarantee the system is a topologically non-trivial topological state in absence of a magnetic field.
It can be seen from Fig. \ref{FIG7}(a) that the band structure has a bandgap with the magnetic flux $\Phi = 0$.
In Fig. \ref{FIG7}(b), the band gap is closed at $k=\pm \pi$ for the magnetic flux $\Phi=N\pi/5$.
As the magnetic flux further increases,
the two bands re-split for the magnetic flux $\Phi=N\pi/2$ shown in Fig. \ref{FIG7}(c), although there is without a global band gap.
And the two bands re-touch at $k=0$ for $\Phi=4N\pi/5$ and reopen again for $\Phi=N\pi$, as plotted in Figs. \ref{FIG7}(d) and \ref{FIG7}(e).
The phase diagram of the system [described in Eq.~(\ref{eq8})] in the plane of Zeeman field $E_z$ and magnetic flux $\Phi$ is shown in Fig.~\ref{FIG7}(f).
One can see from Fig. \ref{FIG7}(f) that the topological invariants is $Z_2=1 (0)$ blue (white) region.
For $E_z=2$,  the system undergoes twice topological phase transitions from $Z_2=1$ to $Z_2=0$ and from $Z_2=0$ to $Z_2=1$ at $\Phi=N\pi/5$ and $\Phi=4N\pi/5$, respectively.
Figures \ref{FIG7}(a) - \ref{FIG7}(e) correspond to the five points (0, 2), ($N\pi/5$, 2), ($N\pi/2$, 2), ($4N\pi/5$, 2) and ($N\pi$, 2) of the corresponding phase diagram plane ($\Phi,E_z$).
In Figs. \ref{FIG7}(a) and \ref{FIG7}(e), the system is the topologically non-trivial phase with topological invariants $Z_2=1$ and is the topologically trivial phase with $Z_2=0$ in Fig. \ref{FIG7}(c).
Figures \ref{FIG7}(b) and \ref{FIG7}(d) show the critical points of topological phase transition in the phase diagram.
These conclusions show the topological phase transitions induced
by the magnetic flux in the topologically non-trivial phase.

\begin{figure}
\includegraphics[width=8.5cm, clip=]{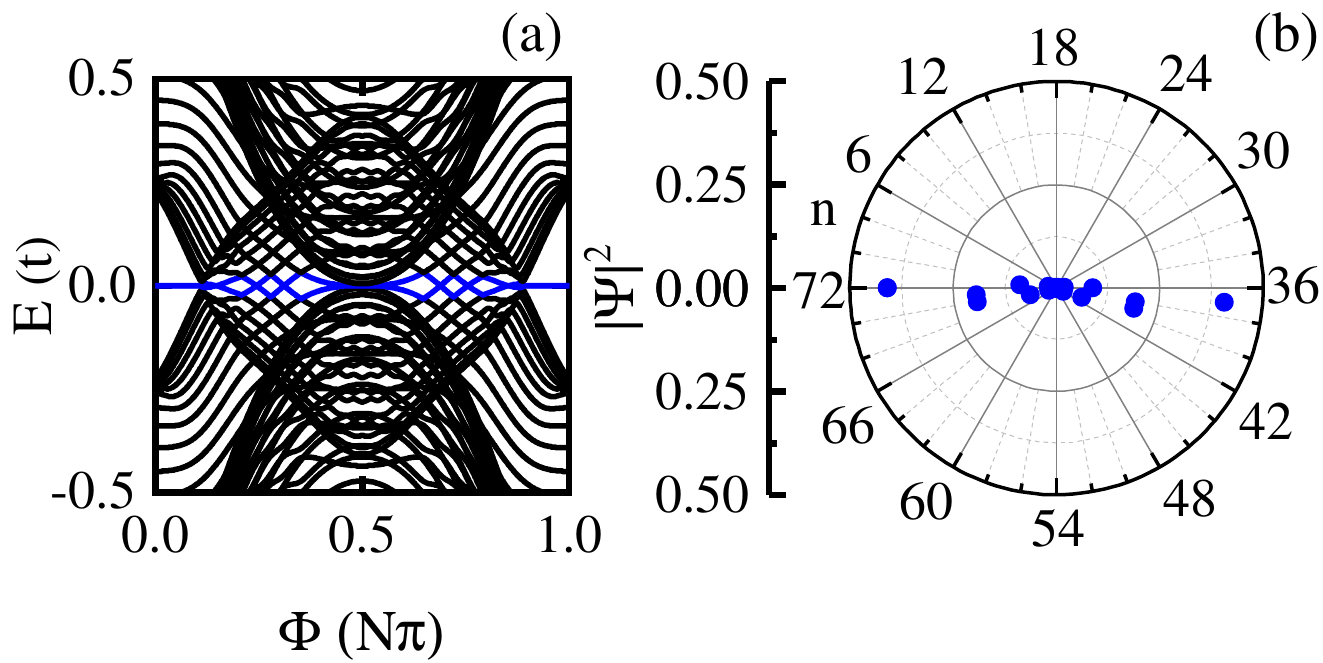}
\caption{(a) The energy spectrum of the Rashba semiconductor nanowire ring as a function of magnetic flux $\Phi$.
Here $E_z=0$ for 1 to 36 sites and $E_z=2$ for 37 to 72 sites.
The flat band so-called MZMs (highlighted by the blue line) emerges as a function of magnetic flux $\Phi$. (b) Polar plot of the
probability distribution of wave function $\left| \Psi \right|^2$ versus the site $n$ with the energy E = 0 for $\Phi=0$.
The elevant parameters are $t=1, \Delta =1,~\mu=0, \alpha=1, N=72$.}
	\label{FIG8}
\end{figure}

To control the creation and annihilation of the MZMs in
the Rashba semiconductor nanowire ring,
we propose a Rashba semiconductor nanowire ring,
one-half of which is always in the topologically trivial phase
and the other half can transit between the topologically non-trivial and trivial phases with the change of the magnetic flux.
According to the phase diagram in Fig. \ref{FIG7}(f),
it can be concluded that the system is always topologically trivial phase at the Zeeman field $E_z=0$ no matter how the magnetic flux changes.
Thus the Zeeman field of one half of our proposed quantum ring is fixed at $E_z=0$ and it of the other half is set to be $E_z=2$.
In Fig. \ref{FIG8}(a), we plot the energy spectrum of our proposed semiconductor nanowire ring as a function of magnetic flux. One can see that the energy spectrum shows the MZMs while the magnetic flux $\Phi$ is less than $\Phi=N\pi/5$ or greater than $\Phi=4N\pi/5$. The reason is that one half of the ring is always topologically trivial phase and the other half part of the
ring is in the topologically non-trivial phase in this case.
However, the MZMs will disappear while a half of the ring undergoes the transition of the topologically non-trivial phase to trivial at $\Phi=N\pi/5$,
as shown in Fig. \ref{FIG7}(f). In Fig. \ref{FIG8}(b) we also display the probability distribution of wave function of zero-energy states as a function of site. One can see that the probability distribution of the zero-energy states is mainly located at the site between two types of the topological phase in real space. It is worth noting that there are some zero-energy points for the magnetic flux
$\Phi$ between $ N\pi/5$ and $4N\pi/5$,
which are not MZMs but the bulk energy mixing induced by the magnetic flux.

\section{\uppercase\expandafter{\romannumeral 4}.summary}
In this work, we investigate the energy spectrum and transport property of a one-dimensional Kitaev quantum ring in a threading magnetic field.
It is demonstrated that the gap of the energy spectrum in the topologically non-trivial phase closes and reopens as the strength of a magnetic field increases, accompanied by topological phase transition induced by a magnetic field.
However, the same situation did not occur in the topologically trivial phase, the gap of the energy spectrum does not close with the increasing of a magnetic field. There is no topological phase transition induced by the magnetic field in the topologically trivial phase. Furthermore, we also investigate the energy spectrum of a one-dimensional quantum ring, one half of which is in the topologically trivial phase and the other half is in the topologically non-trivial phase. It is found that the magnetic field can control the appearance and disappearance of the Majorana zero-energy states in our proposed quantum ring.
We also investigate the transport properties of the Kitaev nanowire ring connected by two semi-infinite leads. It is demonstrated that the resonant peaks of transmission coefficient $T_{\textrm{QT}}$ correspond to the critical points of topological phase transition.
Furthermore, we calculate the energy spectrum and the topological number of a more realistic quantum ring model by adopting a semiconductor nanowire with high SOC, superconducting s-wave pairing and Zeeman splitting,
and prove that our findings are universal.

\begin{acknowledgments}
\section{acknowledgments}
This work was financially supported by the National Natural Science Foundation of China (Grants No. 11474084, No. 12074097, and No. 11921005), Natural Science Foundation of Hebei Province (Grant No. A2020205013), and
National Key R and D Program of China (Grant No. 2017YFA0303301).
\end{acknowledgments}


\end{document}